\documentclass[aps,prl,floatfix,reprint,showpacs,superscriptaddress,groupedaddress]{revtex4-1}
\usepackage{latexsym}
\usepackage{graphicx}
\usepackage{newtxmath}
\usepackage{rotating}
\usepackage[normalem]{ulem}
\usepackage{hyperref}
\usepackage{amsmath,amsfonts}
\usepackage{bm}
\usepackage{color}
\usepackage{textgreek}
\usepackage{soul}  % for \st{}
\usepackage{qcircuit}
\usepackage[greek,english]{babel}

\newcommand{\lco}{La$_{2}$CuO$_4$}

\newcommand{\ket}[1]{\ensuremath{| #1 \rangle}}

\usepackage{color}% use colored text in latex

\newcommand{\lyxmathsym}[1]{\ifmmode\begingroup\def\b@ld{bold}
  \text{\ifx\math@version\b@ld\bfseries\fi#1}\endgroup\else#1\fi}

\usepackage{amsfonts}
\usepackage{mathrsfs}
\usepackage{physics}

\hypersetup{colorlinks=true,urlcolor=blue,linkcolor=blue}

\def\a{\alpha}

\def\b{\beta}
\def\D{\Delta}
\def\d{\delta}
\def\e{\epsilon}
\def\g{\gamma}

\def\h{\hat}

\def\s{\sigma}

\def\w{\omega}

\def\dd{\dagger}

\providecommand{\d}[1]{\color{grey}[#1]\color{black}}

\begin{document}
\title{Krylov variational quantum algorithm for first principles materials simulations} 
\begin{abstract}

\end{abstract}

\author{Fran\c{c}ois Jamet$^{1}$}
\email{francois.jamet@npl.co.uk}
\author{Abhishek Agarwal$^{1}$}
\author{Carla Lupo$^{2,1}$}
\author{Dan E. Browne$^{3}$}
\author{Cedric Weber$^{2}$}
\author{Ivan Rungger$^{1}$}
\email{ivan.rungger@npl.co.uk}
\affiliation{$^1$ National  Physical  Laboratory,  Teddington,  TW11  0LW,  United  Kingdom}
\affiliation{ $^2$King's College London, Theory and Simulation of Condensed Matter,
              The Strand, WC2R 2LS London, UK}
\affiliation{$^3$Department of Physics and Astronomy, University College London, Gower Street, London WC1E 6BT, United Kingdom}

\begin{abstract}
We propose an algorithm to obtain Green’s functions as a continued fraction on quantum computers, which is based on the construction of the Krylov basis using variational quantum algorithms, and included in a Lanczos iterative scheme. 
This allows the integration of quantum algorithms with first principles material science simulations, as we demonstrate within the dynamical mean-field theory (DMFT) framework. DMFT enables quantitative predictions for strongly correlated materials, and relies on the calculation of Green’s functions. On conventional computers the exponential growth of the Hilbert space with the number of orbitals limits DMFT to small systems. Quantum computers open new avenues and can lead to a significant speedup in the computation of expectation values required to obtain the Green’s function.
We apply our Krylov variational quantum algorithm combined with DMFT to the charge transfer insulator \lco~using a quantum computing emulator, and show that with 8 qubits it predicts the correct insulating material properties for the paramagnetic phase. We therefore expect that the method is ideally suited to perform simulations for real materials on near term quantum hardware.
\end{abstract}
\maketitle

The Green's function (GF) is a central quantity to compute observables of interest in materials science simulations, such as the density of states (DOS), optical and electronic conductivities or Raman spectra. It is also at the core of many embedding methods, such as the dynamical mean field theory (DMFT)\cite{dmftrev}. A direct computation of the DMFT GF on quantum computer (QC) has been demonstrated on hardware for the smallest possible 2-site DMFT model system. The simulations were based on the so-called Lehmann decomposition, with the required expansion coefficients computed either using variational quantum algorithms (VQAs) \cite{2019arXiv191004735R} or a real-time evolution \cite{Keen2019}. Computations have also been performed for the 2-site Hubbard model \cite{PhysRevResearch.2.033281,zhu2021calculating}. VQAs are rather noise resilient and can scale to an intermediate number of qubits\cite{cerezo2020variational}, while algorithms based on the time-evolution can in principle scale to larger systems, but generally require lower noise levels. Approaches combining time evolution with VQAs have also been proposed \cite{jaderberg2020minimum,cirstoiu2020variational,otten2019noise,lin2021real,Barratt2021}.
Importantly, all QC DMFT approaches require the computation of the ground state (GS) as first step \cite{Bauer2016,wecker2015solving,jaderberg2020minimum, Keen2019,2019arXiv191004735R}, which is a challenge in itself. Algorithms based on quantum phase estimation in principle allow one to obtain the GS of chemical systems accurately \cite{Berry2018,PhysRevX.6.031045,PhysRevX.8.041015,Bauer2016,wecker2015solving}), but they require a large number of high-fidelity qubits, postponing the time where such algorithms can be implemented on hardware.
Practical calculations on current hardware therefore mostly employ VQAs to obtain the GS.

Iterative approaches form ideal frameworks to obtain scalable methods. For computing GFs the so-called Lanczos continued fraction decomposition method falls in this category \cite{Lin1993}. However, the Hilbert space grows exponentially with the number of particles, restricting the method to few atoms on conventional computers. QCs have the potential to overcome this limitation of Lanczos methods\cite{Motta2019}. This can be achieved for example by using simulated imaginary-time evolution \cite{Motta2019}, Trotter time evolution\cite{PRXQuantum.2.010333} or via quantum approximate counting \cite{baker2020lanczos}. 

The Lanczos method requires the construction of the so called Krylov basis states. In this article we present a Krylov variational quantum algorithm (KVQA) to compute this basis, and include it in a Lanczos iterative scheme to compute the GF. We show that it can be integrated with first principles DMFT calculations for real material systems simulations.
By evaluating the required expectation values on a QC the limitation of the exponential scaling of the Hilbert space is overcome. The scalability of the KVQA is mainly determined by the scalability of the underlying VQAs, which is an active area of research \cite{cerezo2020variational}. The developed method allows to access system sizes required to simulate real materials within DMFT. We demonstrate this on the charge transfer insulator La$_2$CuO$_4$ by applying the algorithm using a quantum computing emulator. Our KVQA-DMFT based algorithm correctly predicts the charge transfer insulating behavior in the paramagnetic phase, while density functional theory (DFT) and perturbative methods give a metallic state. 
 
For a given Hamiltonian $\hat H$ and a generally complex energy $z$, the diagonal elements of the retarded GF, $G_{\alpha \s}(z)$, are given by\cite{dmftrev,mahan}
\begin{align}
  G_{\alpha \s}(z) =& \sum_k \frac{e^{-\b (E_k-E_\mathrm{GS})}}{Z}\left.\big\{ \expval{\hat c_{\alpha \s}\left[z-(\hat H-E_k)\right]^{-1} \hat c_{\alpha \s}^{\dd}}{k}\right. \nonumber\\
        &+ \expval{\hat c^{\dd}_{\alpha \s}\left[z+(\hat H-E_k)\right]^{-1}\hat c_{\alpha \s}}{k} \big\},
  \label{eq:retgGeneral}
\end{align}
where $\ket{k}$ is a normalized eigenstate of $\h H$ with energy $E_k$, $E_\mathrm{GS}$ is the GS energy, $Z$ is the partition function, $Z=\sum_k e^{-\b(E_k-E_\mathrm{GS})}$, $\b$ is the inverse temperature, and $\hat{c}_{\a \s }^\dagger$ ($\hat{c}_{\a \s}$) is the creation (annihilation) operator for a particle with spin $\s$ on an orbital with index $\a$. The GF in Eq. (\ref{eq:retgGeneral}) can be obtained by summing up, with the appropriate prefactors, terms of the general form 
\begin{equation}
   g_\phi(z) = \expval{\left[  z - \hat H\right]^{-1}}{\phi},
    \label{eq:g}
\end{equation}
with $\ket{\phi}$ an arbitrary normalized wavefunction.

The so called Krylov space of $\ket{\phi}$ is the linear space spanned by  $\{\phi,\hat H\phi, \hat H^2\phi,...\}$. Using the Lanczos scheme \cite{Lanczos1950}, an orthogonal basis $\{\chi_0=\phi,\chi_1,\chi_2,...\}$ is constructed, such that $\hat H$ is tridiagonal in this basis. If for an arbitrary integer index $n$ we call the diagonal elements of this tridiagonal matrix $a_n$, and the off-diagonal elements $b_n$, 
Eq. (\ref{eq:g}) can be rewritten as continuous fraction \cite{dmftrev}
\begin{equation}
  g_{\phi}(z) = \frac{1}{z - a_0-\frac{b_1^2}{z - a_1-\frac{b_2^2}{z - \a_2....}}}.
\end{equation}
The $a_n$ and $b_n$ are obtained iteratively, together with the construction of the Krylov basis vectors $\ket{\chi_n}$. 
The first state of the Krylov basis is given by $\ket{\chi_0} = \ket{\phi}$, and the corresponding coefficients by
  $a_0 = \expval{\hat H}{\chi_0}$ and $b_0=0$.
The higher order Krylov states and coefficients are iteratively constructed using the relations
\begin{align}
  b_{n}^2 &=\expval{\hat{H}^2}{\chi_{n-1}}-a_{n-1}^2-b_{n-1}^2
  ,\label{eq:bnp1}\\
  \ket{\chi_{n}} &= \frac{1}{b_{n}}[(\hat H-a_{n-1})\ket{\chi_{n-1}}-b_{n-1}\ket{\chi_{n-2}}],\label{eq:unp1}\\
  a_{n} &=\expval{\hat H}{\chi_{n}}.\label{eq:anp1}
\end{align}
This method is well suited for a QC with a universal gate set, where one can prepare the state $\ket{\chi_{n}}$ from an arbitrary initial state $\ket{0}$ by application of a quantum circuit representing an appropriate unitary operator $\hat U_{n}$, so that $\ket{\chi_{n}}=\hat U_{n}\ket{0}$. Typically $\ket{0}$ is chosen in such a way that the state of each qubit is set to zero. The  real numbers $a_{n}$ and $b_{n}$ can then be evaluated by computing the expectation values in Eqs. (\ref{eq:bnp1}) and (\ref{eq:anp1}) on a QC.
\begin{figure}
    \centering
    \includegraphics[width=0.48\textwidth]{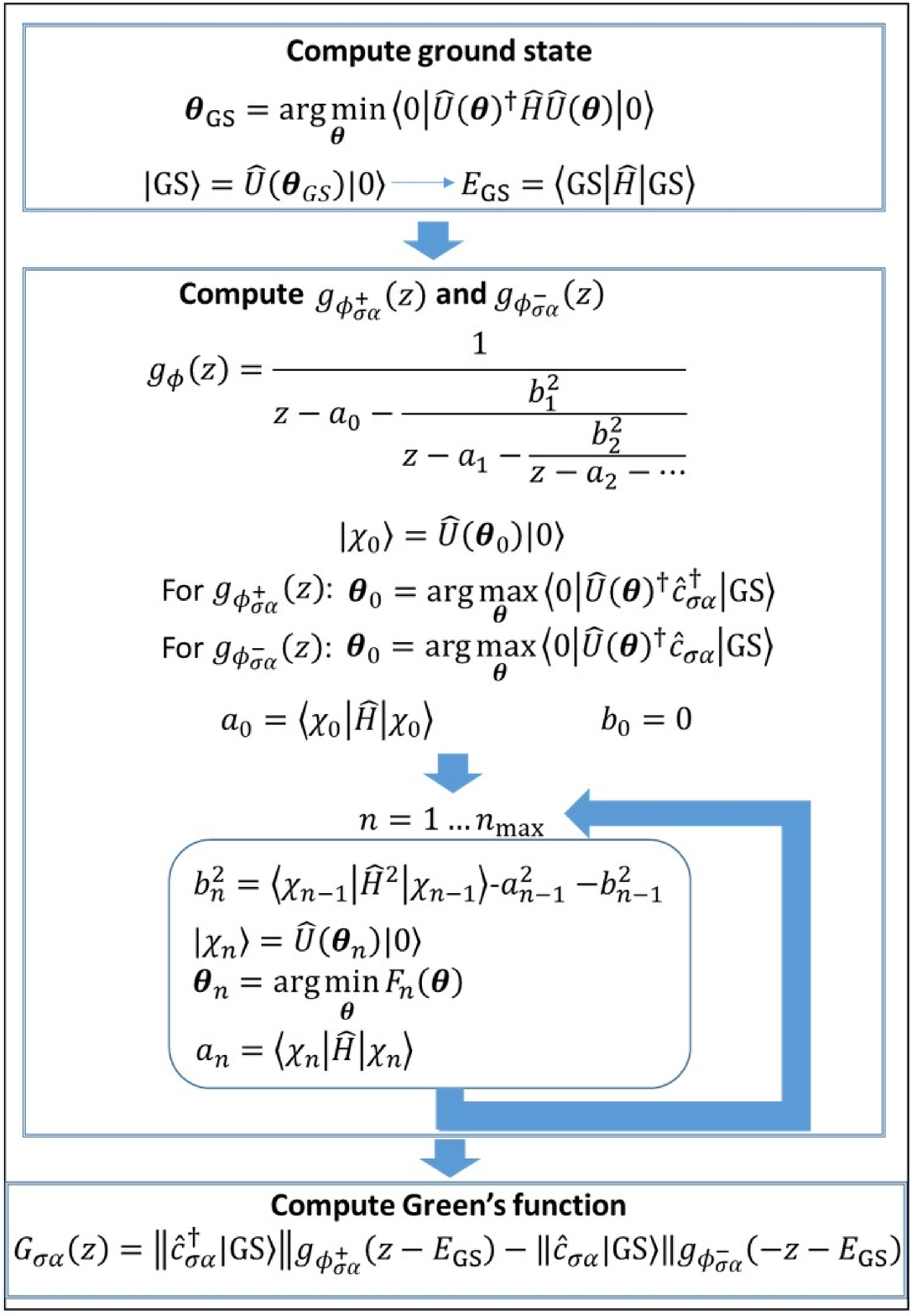}
    \caption{Schematic of the Krylov variational quantum algorithm (KVQA) to compute the Green's function.}
    \label{fig:fig1}
\end{figure}

To map the physical system onto a QC one can use a number of approaches \cite{jordan1928paulische,bravyi2002fermionic,verstraete2005mapping}, and they all result in the creation and annihilation operators being represented by a sum of $N_\gamma$ tensor products of Pauli operators, $\hat P_{\alpha\sigma,j}$, so that $c_{\alpha \sigma}=\sum_j^{N_\gamma} \gamma_{\alpha\sigma,j} \hat P_{\alpha\sigma,j}$ and  $c_{\alpha \sigma}^\dagger=\sum_j^{N_\gamma} \gamma_{\alpha\sigma,j}^* \hat P_{\alpha\sigma,j}^\dagger$, with $\gamma_{\alpha\sigma,j}$ complex numbers. For typical materials simulations $\hat H$ is given by a sum of general products of creation and annihilation operators, so that the transformed form is given by a sum over $N_H$ tensor products of Pauli operators, $P_{j}$, as $\hat H = \sum_{j=1}^{N_H} h_j \hat  P_{j}$, with $h_j$ complex numbers. 

To prepare $\ket{\chi_{n}}$ on a QC we consider unitaries that correspond to the operation of a parametrized quantum circuit, so that $\ket{\chi_{n}}=\hat{U}(\bm{\thetaup}_{n})\ket{0}$, where $\bm{\thetaup}_{n}$ is a vector of real quantum circuit parameters, such as qubit rotation angles.
                    
To obtain $\bm{\thetaup_n}$ we use the relation that if a general wavefunction $\ket{\chi}$ satisfies the three conditions $\matrixel{\chi}{\hat H}{\chi_{n-1}}=b_n$, $\braket{\chi}{\chi_{n-1}}=0$, and $\braket{\chi}{\chi_{n-2}}=0$, then $\braket{\chi}{\chi_n}=1$, so that $\ket{\chi}=\ket{\chi_n}$. This relation can be verified using these conditions and Eq. (\ref{eq:unp1}).
We now define three real positive functions, namely $\epsilon_{n0}(\bm{\thetaup})=\left(\frac{|\bra{0}\hat U^{\dd}(\bm{\thetaup})\h H \hat U(\bm{\thetaup_{n-1}})\ket{0}|}{|b_n|} -1\right)^2$, $\epsilon_{n1}(\bm{\thetaup})=\left|\bra{0}\hat U^{\dd}(\bm{\thetaup}) \hat U(\bm{\thetaup_{n-1}})\ket{0}\right|^2$, and $\epsilon_{n2}(\bm{\thetaup})=\left|\bra{0}\hat U^{\dd}(\bm{\thetaup}) \hat U(\bm{\thetaup_{n-2}})\ket{0}\right|^2$.
Using above relations one obtains that if and only if for a specific vector of angles $\tilde{\bm{\thetaup}}$ we have $\epsilon_{n0}(\tilde{\bm{\thetaup}})=\epsilon_{n1}(\tilde{\bm{\thetaup}})=\epsilon_{n2}(\tilde{\bm{\thetaup}})=0$, then up to a phase shift $\hat{U}(\tilde{\bm{\thetaup}})\ket{0}  = \ket{\chi_{n}}$. To obtain $\bm{\thetaup}_n$ we can therefore concurrently minimize the three functions $\epsilon_{n1/2/3}(\bm{\thetaup})$, which all have a global minimum with a value of zero, and the angles at the minimum correspond to $\bm{\thetaup}_n$.

Equivalently, we can minimize a single cost function, given by
\begin{align}
    F_{n}(\bm{\thetaup})= \sum_{j=0}^{2}w_j \epsilon_{nj}(\bm{\thetaup}),
\end{align}
where we have introduced the positive real-valued weights $w_j$ for each contribution in the sum. For the exact analytical solution $\epsilon_{nj}(\bm{\thetaup})=0$ for all $j$, so that the values of the weights play no role as long as they are finite. In numerical calculations there will always be a small but finite difference from zero in the $\epsilon_{nj}(\bm{\thetaup})$. In this case the values of $w_j$ allow to put more weight on specific terms in the sum. Importantly, since we know that the exact result for each $\epsilon_{nj}(\bm{\thetaup})$ should be zero at the minimum, this enables us to verify the quality of the converged result at each $n$ by checking the difference of the values from zero.
To compute $\epsilon_{n1}(\bm{\thetaup})$ and $\epsilon_{n2}(\bm{\thetaup})$ on a QC we use the method outlined in Ref. \cite{2019arXiv191004735R}. To obtain $\epsilon_{n0}(\bm{\thetaup})$ one might use the Hadamard test protocol \cite{mitarai2019methodology,aharonov2009polynomial}, but in our computations we use a different method, which is optimized taking into account the properties of the Krylov basis, and presented in detail in the Supplementary Material (SM).

We now use the KVQA presented above to obtain the GF in Eq. (\ref{eq:retgGeneral}). Here we consider the zero temperature case with non-degenerate GS, while the general case is presented in the SM. From Eqs. (\ref{eq:retgGeneral}) and (\ref{eq:g}) we obtain the zero temperature diagonal elements of the GF as
\begin{align}
G_{\sigma \a}(z) =& ||\hat c_{\a \s }^{\dd}\ket{\mathrm{GS}}|| \;g_{\phi_{\a \s}^+}(z-E_\mathrm{GS})
\nonumber\\
  &-||\h c_{\a \s}\ket{\mathrm{GS}}|| \;g_{\phi_{\a \s}^-}(-z+E_\mathrm{GS}),
\end{align}
where we have introduced the auxiliary normalized wavefunctions $\ket{\phi_{\a \s}^{+}} {=} \frac{\hat c_{\a \s}^{\dd}\ket{\mathrm{GS}}}{||\hat c_{\a \s}^{\dd}\ket{\mathrm{GS}}||}$ and  $\ket{\phi_{\a \s}^{-}} {=}\frac{\hat c_{\a \s}\ket{\mathrm{GS}}}{||\hat c_{\a \s}\ket{\mathrm{GS}}||}$, with $\ket{\mathrm{GS}}$ the ground state for a given $\hat{H}$. We compute the $g_{\phi_{\a \s}^\pm}$ on a QC using our iterative algorithm presented above with the inital state $\ket{\phi_{\a \s}^\pm}$. The norms $||\hat c_{\a \s}^{\dd}\ket{\mathrm{GS}}||$ and $||\hat c_{\a \s}^{\dd}\ket{\mathrm{GS}}||$ can be computed on a QC in an analogous way to the computation of the expectation value of $\hat{H}$. If $||\hat c_{\a \s}^{\dd}\ket{\mathrm{GS}}||=0$ the corresponding term in the GF is $0$, so that $\ket{\phi_{\a \s}^{+}}$ does not need to be calculated (and analogously for $\ket{\phi_{\a \s}^{-}}$). To obtain $\ket{\mathrm{GS}}$ a number of algorithms can be used \cite{peruzzo2014variational,mcardle2019variational,aspuru2005simulated}. In our application to a real material we use the variational quantum eigensolver (VQE) method \cite{peruzzo2014variational}. As a remark, once $a_n,b_n$ are computed, the GF can be evaluated either for complex energies or also directly for the real axis. This avoids the ill-defined procedure of the analytical continuation, which is required for methods, where it can only be computed on the complex Matsubara frequencies \cite{PhysRevE.94.023303}. The full KVQA is illustrated schematically in Fig. \ref{fig:fig1}.

On a conventional computer, the Lanczos method scales as the size of the Hilbert space, i.e exponentially with the number of orbitals. On the other hand, on a quantum computer the cost of computing term as $\bra{k}\hat H \ket{j}$ is proportional to the number of terms in the Hamiltonian, so that it has a polynomial scaling with respect to the number of orbitals. However, we note that the number of steps required to optimize the quantum circuit unitary $\hat{U}(\bm{\thetaup}_{n})$ at each iteration depends on the optimizer and can become large, and furthermore the potential presence of barren plateaus can make it difficult to find the global minimum. 
The accuracy obtained at each iteration step depends mainly on the quality of the quantum circuit unitary $\hat{U}(\bm{\thetaup}_{n})$, which is affected both by noise in the quantum hardware and by the finite number of parametric gates used. We note that for some applications, rather than computing the GF using the continued fraction expansion, it is advantageous to expand the GF in terms of its so called moments\cite{tilly2021reduced,PhysRevB.103.085131}. For these cases a modified iterative KVQA can be used to construct the Krylov basis, which we present in the SM.

The KVQA allows to perform simulations for real materials systems on a QC. To demonstrate this, we use it to compute the GF of an Anderson impurity model (AIM). The AIM is used for strongly correlated materials, where it is necessary to go beyond perturbative approaches.
For materials simulations the AIM is at the core of the DMFT\cite{dmftrev},
where the correlations of periodic systems are assumed to be purely local,
and the lattice problem is mapped to an AIM. The AIM Hamiltonian is given by
\begin{align}
      \hat H =& \sum_{\substack{\a\b\\ \s\s'}} \mu_{\a \b \s\s'} \hat c_{\alpha\s}^{\dd}\hat c_{\b\s'} + \sum_{\substack{\a\b\g\d\\ \s\s'}} U_{\a\b\g\d} \hat c^{\dd}_{\a \s} c^{\dd}_{\b \s'}\hat c_{\g\s'} \h c_{\d\s} 
 \nonumber \\ &+ \sum_{i\a \s}\left( V_{i\a} \hat f_{i\s}^{\dd}\hat c_{\a\s} +\mathrm{h.c.}\right)
 +  \sum_{ij  \s} \e_{ij} \hat f_{i\s}^{\dd}\hat f_{j\s},
     \label{eq:AIM}
\end{align}
where $\hat c_{\a\s}^{\dd}$ and $\hat c_{\a\s}$ ($\hat f^{\dd}_{i \s},\hat f_{i \s}$)  are the creation and annihilation operators on the impurity sites (bath sites), $\mu_{\a \b \s\s'}$ are the impurity single-particle Hamiltonian elements, $ U_{\a\b\g\d}$ is the impurity interaction tensor, $V_{j\a}$ are the bath-impurity hopping parameters, and $\epsilon_{ij}$ are the bath Hamiltonian elements. 
There are different approaches to solve the AIM \cite{dmftrev,ctqmcrev}. Here we consider the exact diagonalization (ED) method \cite{PhysRevLett.72.1545,Liebsch_2011,dmftrev}, where the bath is discretized by a finite number of bath sites. On conventional computers the exponential scaling of the Hilbert space with the number of sites limits the accessible system sizes. The currently largest computed system has 24 sites (impurity + bath)\cite{PhysRevB.73.205121}, so that it is desirable to develop a method on a QC that might allow  overcoming this limitation.

\begin{figure}
    \centering
    \includegraphics[width=0.48\textwidth]{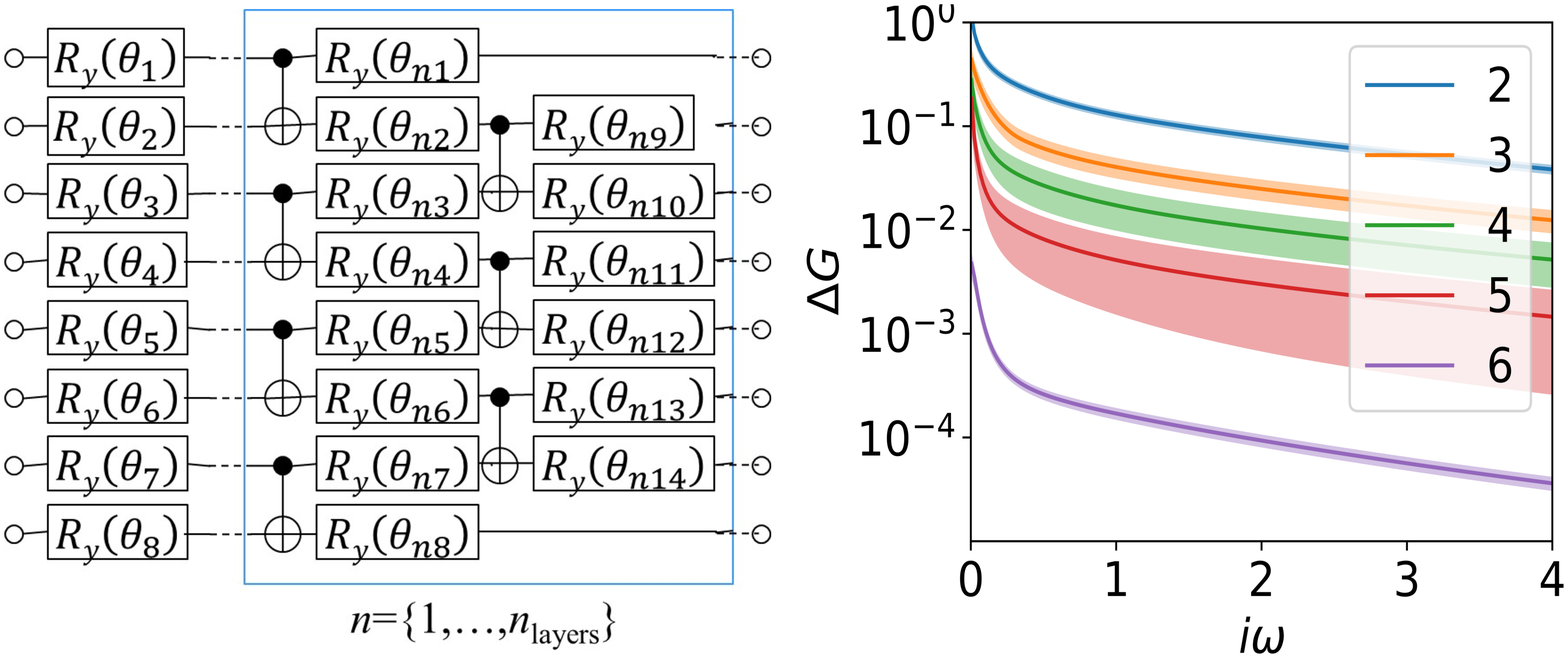}
    \caption{a) State preparation quantum circuit for the 8 qubits AIM; the ansatz can be systematically improved by increasing $n_\mathrm{layers}$.
    b) Relative difference between the Green's function computed on a quantum emulator using the KVQA and on a conventional computer using a classical algorithm ($\D G {=} \frac{|G^\mathrm{KVQA}(i\w) -G^\mathrm{conv}(i\w)|}{G^\mathrm{conv}(i\w)}$), for increasing $n_\mathrm{layers}$ from 2 to 6.  The result is averaged over a set of 80 random parameters in the AIM. The different curves show $\D G $ as a function of the number of layers used for the ansatz. }
    \label{fig:fig2}
\end{figure}
\begin{figure}
    \centering
    \includegraphics[width=0.48\textwidth]{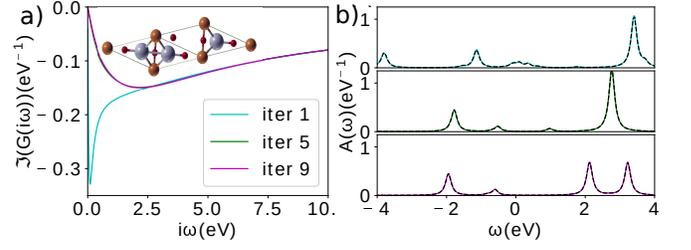}
        \caption{Evolution of the imaginary part of the Green's function on (a) the Matsubara axis and (b) for real energies at different iterations in the DMFT loop for \lco. The impurity DOS is given by $A(\omega) = - \frac{1}{\pi} Im(G(\omega))$. The solid curves are obtained using KVQA on an emulator, and the dashed curves represent the results obtained with a conventional computing algorithm for comparison, showing good agreement.}
    \label{fig:fig3}
  \end{figure}

We demonstrate our Lanczos based KVQA-DMFT method integrated with first principles calculations for \lco~(see the inset of Fig \ref{fig:fig3}a for a representation of the atomic structure). \lco~is a well-known charge transfer correlated insulator, which cannot be appropriately described by DFT, and requires corrections with higher levels of theory, namely the quasiparticle self-consistent \emph{GW} approximation (QS\emph{GW})\cite{PhysRevB.76.165106}  combined with DMFT \cite{Choi2016}. We first perform the QS\emph{GW} simulation using the Questaal package\cite{questaal} for the  orthorhombic phase of \lco~with space group 64/Cmca\cite{PhysRevB.73.144513}. The static QS\emph{GW} self-energy is computed on a 4x4x4 $k$-mesh. 
The DMFT subspace is then constructed via local projection onto the Cu $3d_{x^2-y^2}$ orbital of the Cu augmentation spheres. The AIM to be solved within the DMFT loop then consists of a single impurity Cu $d_{x^2-y^2}$ orbital coupled to the bath. The Hubbard interaction used is $U{=}10$ eV as in previous studies \cite{Choi2016,PhysRevX.8.021038}. For our QC simulations we use 4 sites in total (1 impurity + 3 bath sites). The simulations are run on a QC emulator using the {QuEST} library \cite{Jones2019}, which simulates a noiseless QC with an infinite number of shots. To map the Hamiltonian of Eq. (\ref{eq:AIM}) to a QC we use the Jordan-Wigner transformation \cite{jordan1928paulische,kreula2016non}, which maps the AIM to an 8 qubits system. 
The maximum $n$ in the Lanczos iterative method is at most equal to the size of the Krylov space (see SM). In practice, the continued fraction can be truncated at smaller $n$, since the higher order coefficients contribute progressively less to the DOS. For \lco~the Krylov space is still small enough, so that the largest $n$ in the iterative Lanczos process is chosen to be equal to the size of the Krylov space. 

To obtain accurate results it is important to set up a quantum circuit for the state preparation that is expressive enough to give accurate results for $\ket{\mathrm{GS}}$ and all the considered $\ket{\chi_n}$. A number of state preparation circuits can be used \cite{PRXQuantum.1.020319,cerezo2020variational}, here we choose a hardware efficient ansatz (HEA)\cite{kandala2017hardware}, since it can be efficiently run on the QCs available today. Our HEA consists of alternating layers of $y$-rotations by an arbitrary rotation angle (Ry($\theta$) gates), and CNOT gates (Fig. \ref{fig:fig2}a). The vector of parameters $\bm{\thetaup}$ therefore corresponds to the rotation angles of all the Ry gates. By increasing the number of layers, $n_\mathrm{layers}$, the expressibility of the circuit becomes progressively larger.

During the DMFT loop the GF is computed on the Matsubara imaginary frequencies, since for imaginary energies the GF is usually a smooth function, while on the real axis it has many sharp features. The parameters of $\hat H$ are updated at each iteration, until self-consistency is achieved from one iteration to the next. This requires that the GF can be calculated accurately for all the range that the parameters in $\hat H$ of Eq. (\ref{eq:AIM}) can vary across the different iterations, which is a very stringent test on the stability of the method. We therefore assess the quality of the obtained GF on the Matsubara frequencies for 80 randomly chosen sets of parameters, within the expected range of variation during the DMFT loop [$V{\in}[0,3]$, $\epsilon_{ij}\in[-3,3]$ (with $\e_{ij}{=}\e_{ji}$) , $U{\in}\lbrace{4,8,12\rbrace}$, $\mu{\in}\lbrace{\frac{U}{2}-2,\frac{U}{2},\frac{-U}{2}+2}\rbrace$].

Fig \ref{fig:fig2}b shows the averaged absolute value of the difference between the exact GF and the result obtained with the quantum algorithm for different $n_\mathrm{layers}$, and its standard deviation. The difference is normalized by the absolute value of the GF, and it is important that this relative difference is always small for all energies and parameters. For $n_\mathrm{layers}=2$ the largest relative error is about 0.5, and then progressively decreases for increasing $n_\mathrm{layers}$. For $n_\mathrm{layers}=4$ it is always smaller than 0.02, and it further decreases as $n_\mathrm{layers}$ increases to 6. These results confirm that the GF for such systems is computed accurately with the quantum algorithm over the considered parameter range. To perform the DMFT iteration for \lco~we therefore use a $6$ layers ansatz. The results here indicate that also shorter circuits with $n_\mathrm{layers}=3-4$ can give acceptable results if deeper circuits need to be avoided due to noise in the hardware.

  \begin{figure}
    \centering
    \includegraphics[width=0.4\textwidth]{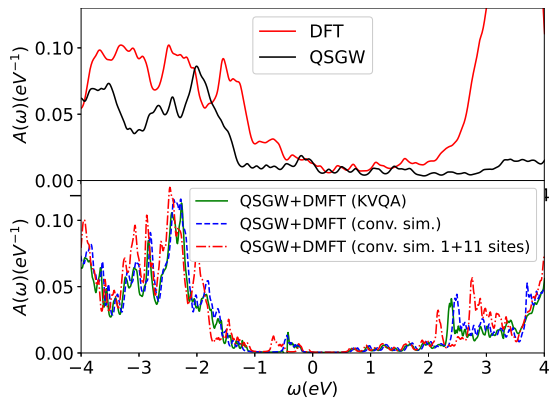}
        \caption{(a)  Total DOS of \lco calculated using paramagnetic DFT
and QSGW. The calculation using DFT results in a large peak at $\omega = 3.5$ eV, originating from the La $f$-states, which in QS\emph{GW} is shifted to higher energy (${\sim}10$ eV). Despite this correction, also QS\emph{GW} still gives an unphysical metallic state. (b) DMFT corrected DOS, obtained both using the KVQA on an emulator (green line), and the conventional computing method (blue dashed line). DMFT correctly predicts an insulating system. As a comparison, the dashed red curve shows the QS\emph{GW}+DMFT DOS  for 11 sites used to represent the bath hybridization, computed with the conventional computing method.}
    \label{fig:fig4}
  \end{figure}
Having verified the general stability of the KVQA to solve the AIM, we can apply it within the DMFT loop for \lco. Fig \ref{fig:fig3} shows the evolution of the GF during the DMFT loop. At each DMFT iteration the AIM is solved on the Matsubara frequencies, and for illustrative purposes also on the real axis, which allows us to observe the evolution of the DOS during the DMFT self-consistency.  In the first iteration the many-body DMFT self-energy is set to zero. Since we start with the paramagnetic QS\emph{GW} state, we obtain a metallic DOS, as can be seen by the finite value of the GF at $\omega=0$. The DMFT loop then progressively converges to the correct insulating state with $A(\omega=0)=0$, and after $9$ iterations it converges to the required accuracy. At each iteration, we also solve the AIM with the conventional computing method, and find excellent agreement with the QC emulator solution (Fig. \ref{fig:fig3}).
Once the DMFT loop is converged we compute the total DOS (Fig \ref{fig:fig4}). Both DFT and QS\emph{GW} give a metallic solution, and only with the addition of DMFT does the paramagnetic \lco~becomes gapped, in agreement with experiments. The DMFT results using the QC algorithm on the emulator are in good agreement with those on conventional computers, both from our own calculations and from literature \cite{Choi2016}. 

To conclude, we have presented and demonstrated the Krylov variational quantum algorithm to obtain the Green's function within first principles DMFT simulations. We chose a variational approach due to its accessibility for near-term quantum computers. Once large scale error-corrected quantum computers become available, improved accuracy might be obtained by adapting the variational components to quantum-simulation-based \cite{Motta2019,PRXQuantum.2.010333} or approximate-counting-based \cite{baker2020lanczos} approaches. Our demonstrations on an emulator showed that our method reproduces very well the Green's functions of the Anderson impurity problem for La$_2$CuO$_4$. This material is an archetypal paramagnetic charge transfer insulator, where non-perturbative beyond-DFT approaches are required, since the local spin fluctuations induce a charge localisation and a
concomitant energy gap in the paramagnetic state. Our results open new avenues for simulating strongly correlated materials, where Mott physics and spin-fluctuations play an important role, on quantum architectures in the near future.

\acknowledgements
FJ, AA, and IR acknowledge the support of the UK government department for Business, Energy and Industrial Strategy through the UK national quantum technologies program. DEB acknowledges funding from UK Engineering and Physical Sciences Research Council (EPSRC) grants EP/T001062/1 and EP/S005021/1. CW was supported by grant EP/R02992X/1 from EPSRC. CL was supported by the EPSRC Centre for Doctoral Training in Cross-Disciplinary Approaches to Non-Equilibrium Systems (CANES, EP/L015854/1). This research project is in part funded by Innovate UK, project Quantifi from UK Research and Innovation, as part of the UK National Quantum Technologies Programme and Industrial Strategy Challenge Fund.

\bibliography{bibli}

\end{document}

\title{Supplementary material: Krylov variational quantum algorithm for first principles materials simulations}

\author{Fran\c{c}ois Jamet$^{1}$}
\email{francois.jamet@npl.co.uk}
\author{Abhishek Agarwal$^{1}$}
\author{Carla Lupo$^{2,1}$}
\author{Dan E. Browne$^{3}$}
\author{Cedric Weber$^{2}$}
\author{Ivan Rungger$^{1}$}
\email{ivan.rungger@npl.co.uk}
\affiliation{$^1$ National  Physical  Laboratory,  Teddington,  TW11  0LW,  United  Kingdom}
\affiliation{ $^2$King's College London, Theory and Simulation of Condensed Matter,
              The Strand, WC2R 2LS London, UK}
\affiliation{$^3$Department of Physics and Astronomy, University College London, Gower Street, London WC1E 6BT, United Kingdom}

\maketitle

\section{Cost function evaluation on a quantum computer}

The different terms of the cost function in Eq. (7) in  the main manuscript need to be computed on the quantum computer (QC) individually. The two terms  $|\bra{0}\hat U(\bm{\thetaup}) \hat U(\bm{\thetaup_{n-1}})\ket{0}|$ and $ |\bra{0}\hat U(\bm{\thetaup}) \hat U(\bm{\thetaup_{n-2}})\ket{0}|$ can be computed on a QC by preparing the initial state $\ket{0}$, then applying the quantum circuit $\hat U(\bm{\thetaup}) \hat U(\bm{\thetaup_{n-1}})$ (resp. $\hat U(\bm{\thetaup}) \hat U(\bm{\thetaup_{n-2}})$), and then measuring the probability to find the state $\ket{0}$. The first term in the sum of the cost function on the other hand requires to compute $|\bra{0}\hat U(\bm{\thetaup})\h H \hat U(\bm{\thetaup_{n-1}})\ket{0}|$. In this section we present two methods to compute this quantity. 

The first method is based on the so-called Hadamard test \cite{mitarai2019methodology,aharonov2009polynomial}. Within this method one uses the fact that $\hat H$, once the Jordan-Wigner transform is performed, is a sum of Pauli strings $\hat H = \sum_i h_i \hat P_i$, so that one needs to compute
\begin{equation}
    |\bra{0}\hat U(\bm{\thetaup})\h H \hat U(\bm{\thetaup_{n-1}})\ket{0}| = |\sum_i h_i \bra{0}\hat U(\bm{\thetaup})\h P_i \hat U(\bm{\thetaup_{n-1}})\ket{0}|.
\end{equation} 
On a quantum computer we can compute each $\bra{0}\hat U(\bm{\thetaup})\h P_i \hat U(\bm{\thetaup_{n-1}})\ket{0}$ individually. Here we note that we need to compute not only the amplitude, but also the phase of this quantity, which we denote as
\begin{equation}
    p_{i,n-1}(\bm{\thetaup}) = \bra{0}\hat U(\bm{\thetaup})^\dagger\h P_i \hat U(\bm{\thetaup_{n-1}})\ket{0}.
\end{equation}
The Hadamard test protocol to  evaluate the $p_{i,n}$ individually on a QC is  described below for completeness:
\begin{enumerate}
    \item Prepare an ancilla qubit in the state $\ket{+}$, and all other qubits in the state $\ket{0}$. To avoid confusion with the state of the ancilla qubit, in describing the Hadamard test protocol we denote the collective states of the non-ancilla qubits using a bold font, e.g. $\ket{\bm{0}}$ represents the state when all non-ancilla qubits are in the state $\ket{0}$.
    \item Apply an ancilla controlled $\hat{U}(\bm{\thetaup})^\dagger \hat{P}_j \hat{U}(\bm{\thetaup}_n)$ gate on the non-ancilla qubits, which prepares the non-ancilla qubits in the state $\ket{\bm{\phi}} = \hat{U}(\bm{\thetaup})^\dagger \hat{P}_j \hat{U}(\bm{\thetaup}_n) \ket{\bm{0}}$ if the ancilla qubit is in the state $\ket{1}$, while it leaves them in the $\ket{\bm{0}}$ state if the ancilla qubit is in the state $\ket{0}$. This results in the state
\begin{equation}
\frac{1}{\sqrt{2}}(\ket{0}\otimes \ket{\textbf{0}} + \ket{1}\otimes \ket{\bm{\phi}}).
\end{equation}

    \item Apply a Hadamard gate on the ancilla qubit to get the state
    \begin{equation}
\frac{1}{2}\Bigl[\ket{0}\otimes (\ket{\bm{0}} + \ket{\bm{\phi}}) + \ket{1}\otimes(\ket{\bm{0}} - \ket{\bm{\phi}})\Bigr].
\end{equation}
    \item Measure and calculate the expectation value of the ancilla qubit in the computational basis, giving
    \begin{equation}
\expval{\hat \sigma_z^{ancilla}} = \frac{1-\Re{\bra{\bm{0}}\ket{\bm{\phi}}}}{2}.
\end{equation}
\item The value $\Im{\bra{\bm{0}}\ket{\bm{\phi}}}$ can be calculated in an analogous way, except that instead of preparing the ancilla qubit in the state $\ket{+}$, we prepare it in the state $\ket{0} + i\ket{1}$. Note that if $p_{j,n-1}$ is known to be real due to, for example, symmetries in the system, then one does not need to calculate $\Im{\bra{\bm{0}}\ket{\bm{\phi}}}$.
\end{enumerate}
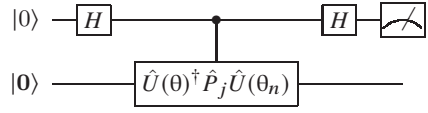
\begin{figure}
\[ \Qcircuit @C=1em @R=1em { 
\lstick{\ket{0}} & 			\gate{H} & \ctrl{1} 		& 	\gate{H} &	\meter\qw \\ 
\lstick{\ket{\textbf{0}}} & \qw & \gate{\hat{U}(\bm{\thetaup})^\dagger \hat{P}_j \hat{U}(\bm{\thetaup}_n)} &\qw 	&		\qw \qw } \]
\caption{Circuit diagram for measuring the real component of $p_{j,n-1}(\bm{\thetaup}) = \bra{\bm{0}}\hat U(\bm{\thetaup})^\dagger\h H \hat U(\bm{\thetaup_{n-1}})\ket{\bm{0}}$ using the Hadamard test method, which requires an ancilla qubit.The first horizontal line corresponds to the ancilla qubit prepared in the initial state $\ket{0}$, and the second horizontal line corresponds to the non ancilla qubits, which are all prepared in the state $\ket{\bm{0}} = \ket{0}\otimes \ket{0}\otimes ...\otimes\ket{0}$. $H$ in the figure represents the Hadamard gate.}
\label{fig_circuit_overlap_measurement}
\end{figure}
The circuit diagram for this protocol is shown in Fig. \ref{fig_circuit_overlap_measurement}. 

We note that the Hadamard test described above requires an ancilla controlled gate, which is  expensive to implement on quantum computers due to the typically lengthy decomposition of the gate into native gates of a device. Hence we developed an alternative method to directly compute $\left|\bra{0}\hat U(\bm{\thetaup})\h H \hat U(\bm{\thetaup_{n-1}})\ket{0}\right|$, which does not require ancilla qubits. It is based on the application of a short time evolution, implemented via a Suzuki-Trotter expansion\cite{Trotter1959,Suzuki1993}. For a small time step $\Delta t$ we can write
\begin{equation}
    |\bra{0}\hat U(\bm{\thetaup})^\dagger  e^{i \h H \D t} \hat U(\bm{\thetaup_{n-1}})\ket{0} |= |\bra{0}\hat U(\bm{\thetaup})^\dagger \hat U(\bm{\thetaup_{n-1}})\ket{0}   + i \D t \bra{0}\hat U(\bm{\thetaup})^\dagger \h H  \hat U(\bm{\thetaup_{n-1}})\ket{0}   + \mathcal{O}(\D t ^2) |.  
\end{equation}
Given that error for a first order Suzuki-Trotter expansion is also $\mathcal{O}(\D t ^2)$\cite{Suzuki1993}, we have
\begin{equation}
    |\bra{0}\hat U(\bm{\thetaup})^\dagger  
    \prod_j e^{i h_j \h P_j \D t} 
    \hat U(\bm{\thetaup_{n-1}})\ket{0} |= |\bra{0}\hat U(\bm{\thetaup})^\dagger \hat U(\bm{\thetaup_{n-1}})\ket{0}   + i \D t \bra{0}\hat U(\bm{\thetaup})^\dagger \h H  \hat U(\bm{\thetaup_{n-1}})\ket{0}   + \mathcal{O}(\D t ^2) |.  
\end{equation}
Here, $\mathcal{O}(\D t ^2)$ therefore contains the errors due to the expansion of the exponential and Suzuki-Trotter expansion. 

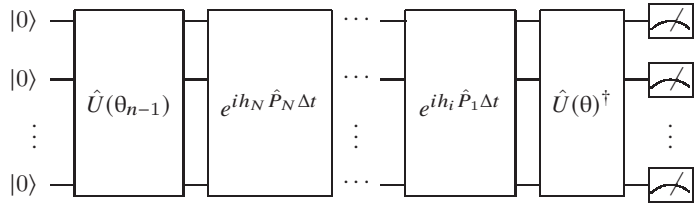
\begin{figure}
\[ \Qcircuit @C=1em @R=1em { 
\lstick{\ket{0}} & \multigate{3}{\hat U(\bm{\thetaup_{n-1}})} & \multigate{3}{ e^{i h_N \h P_N \D t}} & \cdots && \multigate{3}{{ e^{i h_i \h P_1 \D t}}} & \multigate{3}{\hat U(\bm{\thetaup})^\dagger} & \meter\\ 
\lstick{\ket{0}} & \ghost{\hat U(\bm{\thetaup_{n-1}})} & \ghost{{ e^{i h_N \h P_N \D t}}} & \cdots && \ghost{{ e^{i h_1 \h P_1 \D t}}} & \ghost{\hat U(\bm{\thetaup})^\dagger} & \meter\\ 
\lstick{\raisebox{.3em}{\vdots}} & \pureghost{\hat U(\bm{\thetaup_{n-1}})} & \pureghost{{ e^{i h_N \h P_N \D t}}}& \raisebox{.3em}{\vdots} && \pureghost{{ e^{i h_1 \h P_1 \D t}}} & \pureghost{\hat U(\bm{\thetaup})^\dagger} & \raisebox{.3em}{\vdots}\\
\lstick{\ket{0}} & \ghost{\hat U(\bm{\thetaup_{n-1}})} & \ghost{{ e^{i h_N \h P_N \D t}}} & \cdots && \ghost{{ e^{i h_1 \h P_1 \D t}}} & \ghost{\hat U(\bm{\thetaup})^\dagger} & \meter\\ 
 } \]
\caption{Circuit diagram for measuring $|\bra{0}\hat U(\bm{\thetaup})^\dagger  
    \prod_j e^{i \h H_j \D t} 
    \hat U(\bm{\thetaup_{n-1}})\ket{0} |^2$. The QC is initiated to the state where are the qubits are in the state $\ket{0}$. First, we apply the quantum circuit corresponding to the $\h U(\bm{\thetaup_{n-1}})$. Then we applied sequentially the gates corresponding to the exponentiation of the different Pauli strings of the Hamiltonian. Then,  we apply the quantum circuit corresponding to the $\h U^{\dagger}(\bm{\thetaup)}$. Finally, we measure the probability of the QC to be in the state $\ket{0}$.  }
\label{fig_circuit_time_evo_overlap_measurement}
\end{figure}

In the cost function of Eq. (7) in the main manuscript we impose the orthogonality of the basis set, so that at the minimum the term  $\bra{0}\hat U(\bm{\thetaup})^\dagger \hat U(\bm{\thetaup_{n-1}})\ket{0}$ is naught. We can therefore make the following substitution 
in the cost function of Eq. (7) in the main text: 
\begin{align}
|\bra{0}\hat U(\bm{\thetaup})^\dagger \h H  \hat U(\bm{\thetaup_{n-1}})\ket{0}| \rightarrow \frac{1}{\D t}|\bra{0}\hat U(\bm{\thetaup})^\dagger  
    \prod_j e^{i h_j \h P_j \D t} 
    \hat U(\bm{\thetaup_{n-1}})\ket{0}|, 
\end{align}
since the two terms are equal at the minimum of Eq. (7) for $\Delta t \to 0$. We note that the substitution implies that the condition $\bra{0}\hat U(\bm{\thetaup})^\dagger \hat U(\bm{\thetaup_{n-1}})\ket{0}=0$ can be reached for some $\bm{\thetaup}$, which is the case if the circuit ansatz is expressive enough; in practice it will typically be reached at an approximate level.
For a finite $\Delta t$ an additional error of order $\mathcal{O}(\Delta t)$ is introduced.  $|\bra{0}\hat U(\bm{\thetaup})^\dagger  
    \prod_j e^{i h_j \h P_j \D t} 
    \hat U(\bm{\thetaup_{n-1}})\ket{0} |^2$ can be measured on a quantum computer with the circuit shown In Fig. \ref{fig_circuit_time_evo_overlap_measurement}. 
The error  can be reduced by measuring this value  for multiple $\D t$ and extrapolating linearly the value for $\D t = 0$. 

This method does not require ancilla qubits or large circuits conditioned on the ancilla. Also, the number of circuits to run does not increase with the number of terms in the Hamiltonian. We note that with this method the circuit depth scales linearly with the number of terms in the Hamiltonian. Suitable ordering of the Pauli terms in the Hamiltonian can lead to large amount of cancellations, thereby shortening the circuit \cite{gui2020term}. Methods such as circuit recompilation \cite{jaderberg2020minimum} can also be used to reduce circuit depth once the circuit becomes too long. 

\section{Dimension of the Krylov space}
Here, we discuss the size of the Krylov space. In the AIM, the number of particles and the total spin are  conserved quantities. During the iterative Lanczos method, the Krylov basis vectors $\ket{\chi_{n}}$ all have the same total spin $S$, and the total number of particles $N$. We call $N_s$ the total number of sites, i.e impurity+bath sites of the AIM. There are $n_{\ua}=(N + S_z)/2$ (resp. $n_{\da} {=} (N - S_z)/2 $) up (resp. down) electrons. Since there are $\binom{N_s}{n_{\ua}}$ (resp. $\binom{N_s}{n_{\da}}$) combinations of $n_{\ua}$ (resp. $n_{\da}$) electrons that can be put on the $N_s$ sites, the dimension of the subspace is given by
\begin{equation}
    D =  \binom{N_s}{(N+S)/2}\binom{N_s}{(N-S)/2}. 
\end{equation}

\section{Green's function at finite temperature}

The Green's function defined in Eq. (8) in the main text can be generalized to finite temperatures $T>0$ as
\begin{equation}
G^R_{\hat \sigma \alpha}(z) = \sum_n \frac{e^{-\beta (E_n - E_{\mathrm{GS}})}}{Z} \big(||\hat c_{\a \s }^{\dd}\ket{\mathrm{n}}|| \;g_{\phi_{\a \s}^{n+}}(z-E_\mathrm{n}) -||\hat c_{\a \s}\ket{\mathrm{n}}|| \;g_{\phi_{\a \s}^{n-}}(-z-E_\mathrm{n}) \big) ,
  \label{eq:GT}
\end{equation}
where $\ket{n}$ is an arbitrary eigenstate of $\h H$,  $\ket{\mathrm{GS}}$ its ground state,  $E_{\mathrm{GS}}$ is the ground state energy, $Z$ the partition function $Z{=}\sum_n e^{-\b(E_n-E_{\mathrm{GS}})}$, $\b$ the inverse temperature, $\phi_{\a \s}^{n+} {=} \frac{\hat c_{\a \s}^{\dd} \ket{n}}{||\hat c^{\dd}_{\a \s} \ket{n}||}$ and  $\phi_{\a \s}^{n-} {=} \frac{\hat c_{\a \s} \ket{n}}{||\hat c \ket{n}||}$. The Boltzmann factor limits the number of states, which contribute to the sum. In practice, we define a cutoff $\omega_{b}$, and compute the contributions $g_{\phi_{\a \s}^{n-}}$ and $g_{\phi_{\a \s}^{n-}}$ for given $n$ only if $\frac{e^{-\beta (E_n - E_{\mathrm{GS}})}}{Z} {>} \omega_b$. We note that when the temperature increases, the number of excited states, which contribute to the sum in Eq. (\ref{eq:GT}), also increases. In order to get the ground state and the first excited states, a variational quantum eigensolver method  can be performed \cite{Higgott2019}. Once the ground state and the required excited states are computed,  $g_{\phi_{\a \s}^{n-}}$ and $g_{\phi_{\a \s}^{n-}}$ are computed using the Krylov variational quantum algorithm (KVQA) presented in the main text. 

\section{Moments}

Here, we present the algorithm to compute the moments of  the Hamiltonian, which are defined as
\begin{align}
  \mu_n&=\expval{\hat H^n}{\phi}.
  \label{eq:mun}
\end{align}
In principle,  all the $\mu_n$ can be computed on a quantum computer by directly evaluating the expectation values in Eq. (\ref{eq:mun}). For instance, the first moment is simply the energy of the state $\phi$. To reduce the number of terms to compute it is possible to find the tensor product basis \cite{Vallury2020} to reduce the number of measurements. However, when the order of the moments grows, the number of independent vectors in the tensor product basis grows as  $\mathcal{O}(3^{N_q})$, where $N_q$ is the number of qubits. This scaling makes the calculation of the moments on a quantum computer intractable for large $N_q$.
Below we outline a method for computing the moments directly on a quantum computer, similar in spirit to the Krylov basis computation within a Lanczos algorithm outlined in the main manuscript.

After the Jordan-Wigner transformation, the fermionic Hamiltonian can be mapped to a spin Hamiltonian $\hat H {=} \sum_i h_i \hat P_i$, where the $P_i$ are Pauli strings. The first step of the algorithm is to compute the first two moments as
  \begin{align}
    \mu_1&=\expval{\hat H}{\phi}  = \sum_{i=1}^N h_i  \expval{ \hat U_{\phi}^\dagger\hat P_i\hat U_{\phi}}{0} \\
    \mu_2&=\expval{\hat H^2}{\phi}  = \sum_{i=1}^N\sum_{j=1}^N h_i h_j  \expval{\hat U_{\phi}^\dagger\hat P_i\hat P_j\hat U_{\phi}}{0},
  \end{align}
 with $\mu_2\ge 0$.
The terms $\expval{\hat U_{\phi}^\dagger\hat P_i\hat U_{\phi}}{0}$ and $\expval{\hat U_{\phi}^\dagger\hat P_i\hat P_j\hat U_{\phi}}{0}$  can be computed on a QC as expectation value of Pauli strings. If $\ket{\phi}$ is the ground state wave function, then $\mu_1=E_\mathrm{GS}$, where $E_\mathrm{GS}$ is the energy of the ground state.

We can now define the state
\begin{equation}
   \ket{\Phi_2} = \frac{1}{\sqrt{\mu_2}}\hat H\ket{\phi},
\end{equation}
so that $\braket{\Phi_2}{\Phi_2}=1$, and is therefore a state that can be represented on a QC as $\ket{\Phi_2}=\hat U_{\Phi_2}\ket{0}$. 
We can then compute $\mu_3=\mu_2\expval{\hat H}{\Phi_2}$ and $\mu_4=\mu_2\expval{\hat H^2}{\Phi_2}$, and define $\ket{\Phi_4} = \sqrt{\frac{\mu_2}{\mu_4}}\hat H^2\ket{\Phi_2}$. We can iterate this process for progressively higher moments. Note that for the special case, where $\mu_2=0$, one also has that $\mu_1=0$ and all the other moments are also zeros by construction. 

The general iteration procedure therefore goes as follows. Once all moments up to order $2n$ are computed, and with them all the auxiliary states up to the same order, then we can compute the next two moments as
\begin{align}
  \mu_{2n+1}&= \mu_{2n} \expval{\hat H}{\Phi_{2n}},\\
  \mu_{2\left(n+1\right)}&= \mu_{2n} \expval{\hat H^2}{\Phi_{2n}},
\end{align}
and with them we can then compute
\begin{equation}
   \ket{\Phi_{2\left(n+1\right)}} = \sqrt{\frac{\mu_{2\left(n\right)}}{\mu_{2\left(n+1\right)}}}\hat H\ket{\Phi_{2n}},
   \label{eq:chi2np1}
\end{equation}
with $\ket{\Phi_0}=\ket{\phi}$, and $\mu_0=1$. The state for each $n$ can be represented on a quantum computer as $\ket{\Phi_{2\left(n+1\right)}}=\hat U_{\Phi_{2\left(n+1\right)}}\ket{0}$.
To obtain $\hat U_{\Phi_{2\left(n+1\right)}}$ we use a variational quantum approach, where we use an parametrized ansatz $U(\bm{\thetaup})$ as we do in the main manuscript, and we optimize the set of parameters $\theta$ to maximize the overlap
\begin{equation}
\begin{split}
    \left|\bra{\hat U(\bm{\thetaup})}\ket{\Phi_{2\left(n+1\right)}}\right| & = \sqrt{\frac{\mu_{2\left(n\right)}}{\mu_{2\left(n+1\right)}}} \left|\bra{0}\hat U(\bm{\thetaup})\hat H\hat U_{\Phi_{2n}}\ket{0}\right| \\
    & = \sqrt{\frac{\mu_{2\left(n\right)}}{\mu_{2\left(n+1\right)}}} \left|\sum_i h_i \bra{0}\hat U(\bm{\thetaup})\hat P_i\hat U_{\Phi_{2n}}\ket{0}\right|,
\end{split}
\end{equation}
where $\bra{0}\hat U(\bm{\thetaup})\hat P_i\hat U_{\Phi_{2n}}\ket{0}$ is evaluated on a quantum computer.
This procedure can then be iterated to any order.

Computing the moments can be useful in different contexts, such as for computing the ground state energy \cite{Vallury2020}, for methods taking into account dynamical excitations such as EwDMET\cite{PhysRevB.98.235132}, or to get to compute the high frequency expansion of the Green's function, which for instance is used to computed the forces \cite{plekhanov2021calculating,PhysRevB.76.235101}.

\begin{figure}
    \centering
    \includegraphics[scale=0.5]{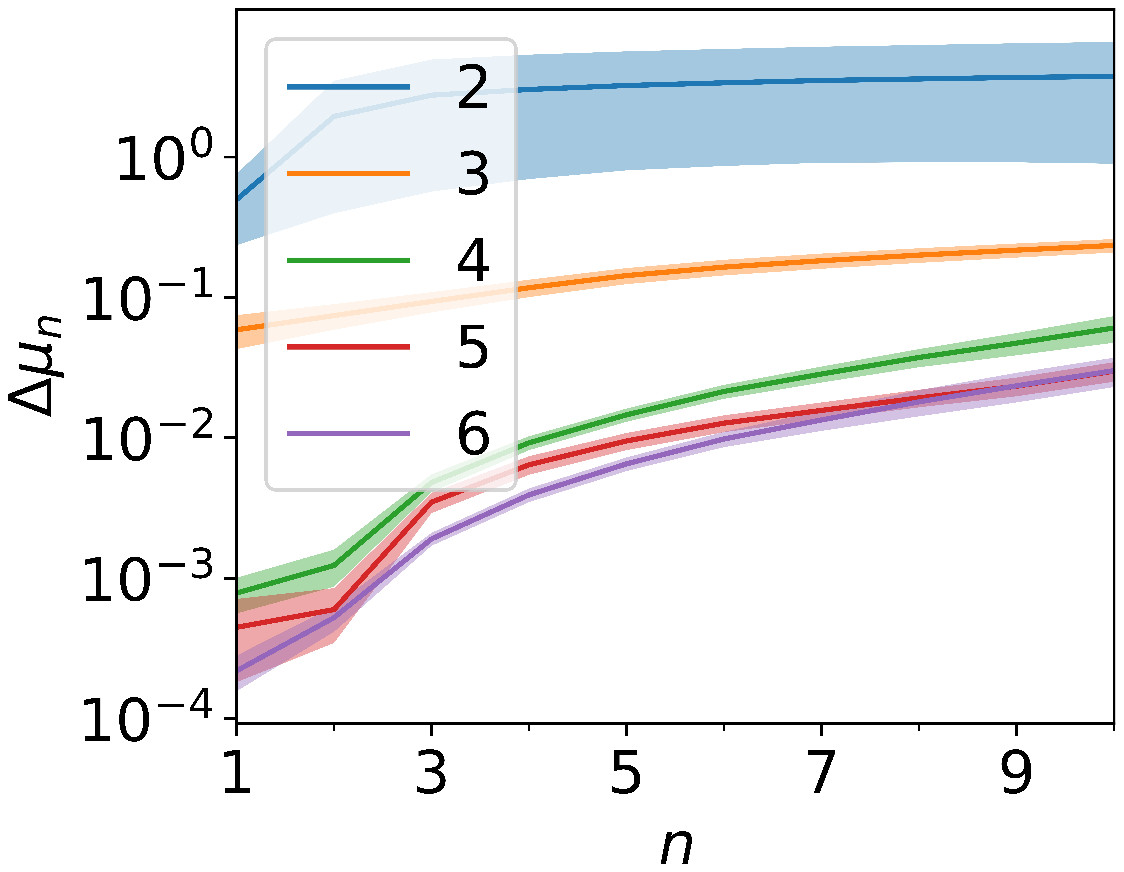}
    \caption{ Average over a set of Anderson impurity model relative error of the moments computed on a  quantum simulator. The  shadows indicate the standard deviation over the different parameters}
    \label{fig:moments}
\end{figure}

 As a representative example, we use this algorithm to compute the moments of the Hamiltonian in Eq. (9) of the main manuscript. Fig. \ref{fig:moments} shows the evolution of the relative error of the moments computed with the method outlined above. The values showed are averaged over 80 sets of parameters, chosen in the same range as for Fig. 2b in the main text. For each set of parameters, we compute the moment up to order $10$. We repeat the calculation for different numbers of layers for the circuit ansatz shown in Fig. 2a. When the number of layers is higher than $4$, the relative error for the first moment is lower than $0.1\%$, and less than $3\%$ for the $\nth{10}$ moment. 

  \section{Relation between the $\mu_n$ and the $a_n, b_n$ coeefficients}
  \label{sec:moments_lanczos}

  If the $a_n$ and $b_n$ coefficients, defined in Eq. (4)-(6) in the main manuscript, are already computed, the moments can also be obtained from those. In this section we therefore show how to compute $\mu_n=\expval{H^n}{\phi}$ when the $a_n$ and $b_n$ are available. For example, $a_n$ and $b_n$ can be constructed using the KVQA presented in the main text. First, we compute $a_n$ and $b_n$ with the Krylov basis, where $\phi$ is the initial state.
  In this Krylov basis, the Hamiltonian matrix $H_{nm}=\bra{\chi_n}\hat{H}\ket{\chi_m}$  can then be written as a tridiagonal matrix
  \begin{equation}
  \bar{H} = \begin{pmatrix}
    a_0 & b_1 & &        & \\
    b_1 & a_2 & b_2& & \\
    &    \ddots & \ddots &\ddots &\\
    &&b_n&a_{n+1}\\
  \end{pmatrix}. \label{eq:H-tridiag}
  \end{equation}
  We then obtain $\mu_n$ as the first element of the matrix resulting by taking the power $n$ of  Eq. (\ref{eq:H-tridiag}). 

\bibliography{bibli}
%\printbibliography